\documentstyle[preprint,aps,epsfig]{revtex}
         
\def\be{\begin{equation}}
\def\ee{\end{equation}}         
\def\ep{\epsilon}
\def\et{\tilde{\epsilon}}

\begin{document}
\draft

\title{Nonlinear alternating current responses of \\ electrorheological solids}
\author{J. P. Huang\footnote{Electronic mail: jphuang@alumni.cuhk.net }}
\address{Department of Physics, The Chinese University of Hong Kong, Shatin, New Territories, Hong Kong and 
Max Planck Institute for Polymer Research, Ackermannweg 10, 55128, Mainz, Germany }

\maketitle

\begin{abstract}

When a composite containing nonlinear dielectric particles suspended in a host medium
is subjected to a sinusoidal  alternating
current (ac)   electric field, the dielectric
response of the composite will generally consist of ac fields at frequencies of higher-order harmonics. For
an electrorheological (ER) solid under structure
transformations due to external fields, we apply the
Ewald-Kornfeld formulation to derive the local electric fields and induced dipole moments explicitly, and then we perform the perturbation expansion method to
extract their fundamental and third-order harmonics analytically. It is shown that the degree of anisotropy of  the
ER solid can affect these harmonics significantly. Our results
are well understood in the spectral representation theory.
Thus, it seems possible to perform a real-time monitoring of the
structure transformation by measuring the nonlinear ac responses of ER solids.
\end{abstract}

\newpage

\section{introduction}

Electrorheological (ER) fluids~\cite{Halsey,Kling98,ShengNM03,PRE-self} have received much
attention due to the potential application of the rapid
field-induced aggregation~\cite{Kling98} and the large
anisotropy~\cite{PRE1}, and they are actually a suspension in
which the induced dipole moment can order the suspended
polarizable particles into columns under the application of a
strong electric field.  Tao and Jiang  examined the temporal evolution of
three-dimensional structure in ER fluids by using a computer simulation~\cite{TaoPRL94}. 
For a wide range of the ratio of the Brownian force to the dipolar force, the ER 
fluid was found~\cite{TaoPRL94} to have a rapid chain formation followed by aggregation of chains into thick columns which has a
body-centered tetragonal (bct) lattice structure, and the Peierls-Landau instability of single chains helps formation of thick
columns. 

 An ER fluid can turn into an ER
solid as the external field exceeds a critical field. For the ER
solid, it is known that its ground state is a bct lattice. One~\cite{Tao98}  proposed
that a structure transformation of the ER solid from the bct
lattice to some other lattices can appear when a magnetic field is
simultaneously applied perpendicular to the electric field and the
particles have magnetic dipole moments. Then, this proposal was
verified experimentally and a structure transformation from the
bct lattice to the face-centered cubic (fcc) lattice was observed
indeed~\cite{Sheng00}. Recently, Lo and Yu have also shown that an
alternative structure transformation from the bct structure to the
fcc can appear under the application of electric fields
only~\cite{Lo01}.

Finite-frequency responses of nonlinear dielectric
composite materials have also attracted much attention both in
research and industrial applications during the last two
decades~\cite{Bergman92}. In particular, when a composite containing
nonlinear dielectric particles embedded in a linear (or nonlinear) dielectric host
medium is subjected to a sinusoidal  alternating
current (ac)  electric field, the dielectric
response of the composite will, in general, consist of ac fields at frequencies of higher-order
harmonics~\cite{PRE1,Levy95,Hui97,Hui98,Gu00,JAP1,WeiJAP}. In experiments, a
convenient method of probing the nonlinear characteristics of the
composite is to measure the harmonics of the nonlinear
polarization under the application of a sinusoidal electric
field~\cite{Kling98}. In this case,  the strength of the
nonlinear polarization should be reflected in the magnitude of the
harmonics. From the theoretical point of view,  the perturbation
expansion~\cite{PRE1,Gu00,WeiJAP} and self-consistent
methods~\cite{PRE1,Wan01} can be used  for extracting such kind of harmonics.

In this work, to investigate the lattice effect  on
the nonlinear ac responses, we shall use the Ewald-Kornfeld
formulation~\cite{Lo01,Ewald,Korn,Lo01-1} to derive the local electric fields and induced dipole moments in the
ER solid which is subject to a structure transformation due to
 external fields, and then perform the perturbation expansion method
to obtain their fundamental and third-order harmonics.  To this end, it is shown that the fundamental and
third-order harmonics are sensitive to the degree of
anisotropy of  ER solids. Thus, by measuring the nonlinear ac responses,
it seems possible to perform a real-time monitoring of the structure
transformation of  ER solids.

This paper is organized as follows. In Sec.~II, we apply the
Ewald-Kornfeld formulation~\cite{Lo01,Ewald,Korn} to derive the local electric field and induced dipole moment in ER solids, and then perform the perturbation expansion
method to extract their fundamental and third-order harmonics. In Sec.~III, we
numerically investigate these harmonics under different conditions, which is followed by a
discussion and conclusion in Sec.~IV.

\section{Formalism}

\subsection{Local electric field and induced dipole moment: Ewald-Kornfeld formulation}

Let us start by considering the ground state of an ER  solid, namely a bct (body-centered tetragonal) lattice, which can be regarded as a
tetragonal lattice, plus a basis of two particles each of which is fixed with a point dipole at its center. One of the two particles is located at a corner and
the other one at the body center of the tetragonal unit cell. Its
lattice constants are denoted by $a(=b)=\ell q^{-1/2}$ and  $c=q\ell$   along the  $x(y)$ and $z$ axes, respectively. In this case, the
uniaxial anisotropic axis is directed along the $z$ axis.
 As $q$ varies, the volume of the unit cell keeps the same, $V_c=\ell^3 .$ Consequently, the degree of anisotropy of the tetragonal lattice is measured by how $q$ is deviated from unity.

In case of an $x-$directed external electric field $\vec{E}_0 ,$ the dipole moments
$\vec{p}=p\hat{x}$ are perpendicular to the uniaxial anisotropic axis.
Then, the local field
$\vec{E}$ (e.g., $\vec{E}=E_x\hat{x}$, $E_z=0$) at the lattice point $\vec{R}=\vec{0}$
has the following form~\cite{Lo01,Ewald,Korn,Lo01-1}

\be
E_x = p\sum_{j=1}^2\sum_{\vec{R}\ne \vec{0}}[-\gamma_1(R_j)+x_j^2q^2\gamma_2(R_j)]-\frac{4\pi p}{V_c}\sum_{\vec{G}\ne \vec{0}}\Pi(\vec{G})\frac{G_x^2}{G^2}\exp(\frac{-G^2}{4\eta^2})+\frac{4p\eta^3}{3\sqrt{\pi}}\label{Ex}.
\ee
In this equation, $\gamma_1$ and $\gamma_2$ are two coefficients, given by

\begin{eqnarray}
\gamma_1(r)&=&\frac{{\rm erfc}(\eta r)}{r^3}+\frac{2\eta}{\sqrt{\pi}r^2}\exp(-\eta^2r^2),\nonumber\\
\gamma_2(r)&=&\frac{3{\rm erfc}(\eta r)}{r^5}+(\frac{4\eta^3}{\sqrt{\pi}r^2}+\frac{6\eta}{\sqrt{\pi}r^4})\exp(-\eta^2r^2),\nonumber
\end{eqnarray}
where ${\rm erfc}(\eta r)$ is the complementary error function, and $\eta$ an adjustable parameter making the summation
converge rapidly. In Eq.~(\ref{Ex}),
$R$ and $G$ denote the lattice vector and the reciprocal lattice vector,  respectively,
\begin{eqnarray}
\vec{R}&=&\ell (q^{-1/2}l\hat{x}+q^{-1/2}m\hat{y}+qn\hat{z}),\nonumber\\
\vec{G}&=&\frac{2\pi}{\ell}(q^{1/2}u\hat{x}+q^{1/2}v\hat{y}+q^{-1}w\hat{z}),\nonumber
\end{eqnarray}
where $l,m,n,u,v,w$ are integers. In addition,  $x_j$ and $R_j$  of Eq.~(\ref{Ex})  are given by, respectively,
$$
x_j=l-\frac{j-1}{2},\,\, R_j=|\vec{R}-\frac{j-1}{2}(a\hat{x}+a\hat{y}+c\hat{z})|,
$$
and the structure factor $\Pi(\vec{G})=1+\exp[i(u+v+w)/\pi]$.

So far, based on a self-consistent method, we apply the result of the local field to evaluate the effective polarizability $\alpha_{eff}$
of the dipole lattice, 
\be
\alpha_{eff}=\frac{\alpha}{1-\alpha\gamma_x/V_c},\label{alpha_eff}
\label{alpha}
\ee
where $\alpha$ stands for the polarizability of an isolated dipole, and $\gamma_x=E_xV_c/p$  the local field factor which was measured in the computer simulations by Martin {\it et al.}~\cite{Mar1,Mar2}.
Let us use $\gamma_z$ and $\gamma_{x}$ ($\equiv\gamma_{y}$) to denote the local-field factors parallel and perpendicular
 to the uniaxial anisotropic axis, respectively. Accordingly, we have $\gamma_z=\gamma_{x}=4\pi/3$ for the bcc lattice ($q=1$).
 In what follows, we set $\gamma'=3\gamma/4\pi$.
 It is worth noting  that $\gamma'$ is a function of a single variable, namely degree of anisotropy $q .$ Also, $\gamma_z'$ and $\gamma_{x}'$  satisfy the sum rule $\gamma_z'+2\gamma_{x}'=3$~\cite{Landau}, and $\gamma_z'=\gamma_{x}'=1$ [at $q=1$ (bcc)] just represents the isotropic limit.

Based on Eq.~(\ref{alpha_eff}), it is straightforward to derive the dipole factor (also called Clausius-Mossotti factor) for a specific particle $b_x'$ and $b_z'$ along the $x$ (or $y$) and $z$ axes, respectively
 \be
 b_{x(z)}'=\frac{b}{1-b\rho \gamma'_{x(z)}},\label{bxin}
 \ee
where $\rho$ stands for the volume fraction of the particles, and $b$ the known dipole factor for an isolated particle, $b=(\epsilon_1-\epsilon_2)/(\epsilon_1+2\epsilon_2)$. Here $\epsilon_1$ and $\epsilon_2$ denote the dielectric constants of the particle and  host fluid, respectively.
 The two parameters, $b_x'$ and $b_z'$,   have been represented as $b_{x(z)}'$ by using the joint subscript $x(z)$. The similar notation are also used for $\gamma_{x(z)}'$. In what follows, for convenience, ${\mathcal X}^{(T)[(L)]}$ will be used to stand for ${\mathcal X}^{(T)}$ and ${\mathcal X}^{(L)}$ which represent the transverse  [i.e., the external field is parallel to the $x$ (or $y$) axis] and longitudinal [namely, the external field is parallel to the $z$ axis]  field cases of the quantity ${\mathcal X}$, respectively.

So far, the local-field effect arising from all the other particles in the lattice has been explicitly included in Eq.~(\ref{bxin}) by using the Ewald-Kornfeld formulation.
Next, let us see the particle as the one embedded in an effective medium with an effective  dielectric constant $\ep_e^{(T)[(L)]}$ which can be determined by
\be
b_{x(z)}' = \frac{\ep_1-\ep_{e}^{(T)[(L)]}}{\ep_1+2\ep_{e}^{(T)[(L)]}}.\label{b'}
\ee
 It is worth remarking that this $\ep_{e}^{(T)[(L)]}$ has included the detailed structural information of the lattice, as expected.

As the external field ${\vec E}_0$ is along the $x$ or $z$ axis, the volume
average of the  local electric field  is given by 
\be 
\langle
{\vec
E}_1^{(T)[(L)]}\rangle=\frac{3\ep_{e}^{(T)[(L)]}}{\ep_1+2\ep_{e}^{(T)[(L)]}}{\vec
E}_0,\label{ET} 
\ee 
where $\langle\cdots\rangle$ stands for the volume average of $\cdots$. In view of
Eqs.~(\ref{bxin})~and~(\ref{b'}), $\ep_{e}^{(T)[(L)]}$ of Eq.~(\ref{ET}) is given by \be
\ep_{e}^{(T)[(L)]}=\frac{b\rho\gamma'_{x(z)}+b-1}{b\rho\gamma'_{x(z)}-2b-1}\ep_1.
\ee 

Then, it takes one step forward to express the corresponding induced dipole moment inside the particle
\be
{\vec p}_1^{(T)[(L)]} = \epsilon_e^{(T)[(L)]}r_a^3b'_{x(z)}{\vec E}_0.\label{PT}\\
\ee


So far, we have derived the local electric fields [Eq.~(\ref{ET})] and induced dipole moments [Eq.~(\ref{PT})], by taking into account the lattice effect in a rigorous manner. In what follows, we shall use $p_1$ (or $\tilde{p}_1$), $\gamma'$, $E_1$ (or $\tilde{E}_1$) and $\ep_e$ to denote both
the transverse and longitudinal field cases, if there are no special instructions.

\subsection{Nonlinear polarization and its higher-order harmonics}

For studying the nonlinear ac responses, let us consider the particle with a cubic nonlinearity like
\begin{equation}
\et_1=\ep_1+\chi_1 E_1^2\approx \ep_1+\chi_1\langle E_1{}^2\rangle ,
\end{equation}
where $\chi_1$ represents the nonlinear susceptibility of the particles embedded in a linear host fluid.  Throughout the paper, we  focus on the weak nonlinearity only, namely $\chi_1\langle E_1{}^2\rangle \ll \epsilon_1$, and the low concentration limit, and we use ``$\sim$'' to indicate the quantities having a nonlinear characteristic.

If a sinusoidal electric field
\be
E_0(t)=E_0\sin\omega t\label{E0t}
\ee
 is applied, the local electric field $\langle \tilde{E}_1\rangle$ and the induced dipole moments $\tilde{p}_1$
will depend on time sinusoidally, too.

Owing to the inversion symmetry of the system, the
 local electric field is a superposition of odd-order harmonics such that~\cite{PRE1,Gu00}
\be
\langle \tilde{E}_1\rangle=E_{\omega}\sin (\omega t)+E_{3\omega}\sin (3\omega t)+\cdots.
\ee
Accordingly, the induced dipole moment contains harmonics as~\cite{PRE1,Gu00}
\be
\tilde{p}_1=p_{\omega}\sin (\omega t)+p_{3\omega}\sin (3\omega t)+\cdots.
\ee
For the purpose of normalization, we shall use $p_0=\ep_2r_a^3bE_0$ which is independent of the nonlinear characteristic as well as the degree of anisotropy $\gamma'$.

\subsection{Analytic solutions}

In what follows, we  apply the perturbation expansion method to extract the harmonics of the local electric fields ($E_{\omega}$ and $E_{3\omega}$) and the induced dipole moments ($p_{\omega}$ and $p_{3\omega}$).

Let us denote the volume average of the linear local electric field 
(namely, the local electric field at which the nonlinear characteristics of the particles disappears)
 [Eq.~(\ref{ET})] as
\be
\langle E_1\rangle\equiv F(\ep_1,\ep_2,\rho,\gamma',E_0).
\ee
In view of the existence of nonlinearity, we have
\be
\langle \tilde{E}_1\rangle\equiv F(\tilde{\epsilon}_1,\ep_2,\rho,\gamma',E_0).
\ee

Next, expand the local field $\langle \tilde{E}_1\rangle$ into a Taylor
expansion, taking $\chi_1 \langle E_1^2\rangle$ as the
perturbative quantity. As a result, we obtain~\cite{PRE1}
 \be \langle \tilde{E}_1\rangle  =
F(\ep_1,\ep_2,\rho,\gamma',E_0)+\frac{\partial}{\partial\et_1}F(\et_1,\ep_2,\rho,\gamma',E_0)|_{\et_1=\ep_1}\chi_1\langle
E_1^2\rangle +\cdots.\label{Taylor-E} \ee
 Keeping the lowest orders of  $\chi_1 E_0(t)^2$ and  $\chi_1 \langle E_1^2\rangle$, because of the time-dependence of the external electric field [Eq.~(\ref{E0t})], we obtain the local electric fields and induced dipole moments in terms of the harmonics, $E_{\omega}$, $E_{3\omega}$, $p_{\omega}$, and $p_{3\omega}$
[note the higher-order harmonics (namely, fifth-, seventh-order, et al.) have been neglected].
Here the harmonics of the local electric fields and induced dipole moments are respectively given by
\begin{eqnarray}
\chi_1^{1/2}E_{\omega} &=& \frac{W_1}{W_2}(\chi_1^{1/2}E_0)-\frac{9\ep_2W_1^2}{4W_2^4}(\chi_1^{1/2}E_0^3)^3, \\
\chi_1^{1/2}E_{3\omega} &=&\frac{3\ep_2W_1^2}{4W_2^4} (\chi_1^{1/2}E_0)^3, \\
p_{\omega}/p_0 &=& -\frac{W_1(\ep_1+2\ep_2)(\Omega_1+\Omega_2)}{4W_2^2\ep_2(\ep_1-\ep_2)},\\
p_{3\omega}/p_0 &=&
\frac{W_1^2(\ep_1+2\ep_2)}{4\ep_2W_2^4(\ep_1-\ep_2)[\gamma'\ep_2\rho+\ep_1(3-\gamma'\rho)]^2}
\sum_{N=1}^5Q_N(\chi_1^{1/2}E_0)^2,
\end{eqnarray}
where
\begin{eqnarray}
W_1 &=& -\gamma'\ep_1\rho+\ep_2(3+\gamma'\rho),\nonumber\\
W_2 &=& \ep_1-\gamma'\ep_1\rho+\ep_2(2+\gamma'\rho).\nonumber \\
\Omega_1 &=&
-\frac{4\ep_1W_2(\ep_1-\ep_2)}{\gamma'\ep_2\rho+\ep_1(3-\gamma'\rho)},\nonumber\\
\Omega_2 &=&
\frac{3W_1}{[\gamma'\ep_2^2\rho(2+\gamma'\rho)+U_1+U_2]^2}\sum_{N=1}^5Q_N(\chi_1^{1/2}E_0)^2,\nonumber
\end{eqnarray}
with
\begin{eqnarray}
Q_1 &=& 4\gamma'\ep_1^3\ep_2\rho(3+\gamma'\rho-\gamma'^2\rho^2),\nonumber\\
Q_2 &=& \gamma'\ep_1^4\rho(3-4\gamma'\rho+\gamma'^2\rho^2),\nonumber\\
Q_3 &=& \gamma'\ep_2^4\rho(6+5\gamma'\rho+\gamma'^2\rho^2),\nonumber\\
Q_4 &=& -2\gamma'\ep_1\ep_2^3\rho(6+7\gamma'\rho+2\gamma'^2\rho^2),\nonumber\\
Q_5 &=&
3\ep_1^2\ep_2^2(-9-3\gamma'\rho+3\gamma'^2\rho^2+2\gamma'^3\rho^3),\nonumber\\
U_1 &=& Q_1/(2\gamma'\ep_1^2\rho),\,\, U_2 =
Q_2/(\gamma'\ep_1^2\rho).\nonumber
\end{eqnarray}

\section{numerical results}

We are now in a position to do some numerical calculations to
discuss the effect of the degree of anisotropy $q$ on the
harmonics of the local electric fields and induced dipole moments.
Without loss of generality, take $\ep_1=30\ep_0$,
$\ep_2=2.25\ep_0$ (dielectric constant of silicone oil), and $\rho=0.1$ for numerical calculations,
where $\ep_0$ denotes the dielectric constant of free space.
In fact, here $\rho = 0.1$  corresponds to the dilute limit.
Nevertheless, it would be realizable as the particles can have a (solid)
hard core and a relatively soft coating (by long polymer chains, etc.)
to avoid aggregation. Thus the model of a soft sphere with an embedded
point dipole can be used. In this case, the many-body (local field)
effects are important, while the multipolar effects can be neglected.
This is exactly part of the emphasis in the present work.
On the other hand, for a higher $\rho$, similar results can be achieved indeed, as expected.

In Fig.~1, we investigate the dependence of the local-field factor\cite{PRE-self}
on the degree of anisotropy $q$. Increasing $q$ causes
$\gamma_{x}'$ (or $\gamma_z')$  to increase (or decrease). Also, a
plateau is shown at $\gamma_{x}'=\gamma_z'=1$, which actually
includes the transformations ranging from the bcc ($q=1$) lattice
to the fcc ($q=2^{1/3}$). Accordingly, similar plateau occurs at
all other figures (see Figs.~2~and~3).

Figure~2 displays the fundamental and third-order harmonics of the
local electric field and induced dipole moment, as a function of
the degree of anisotropy $q$ for different $\chi_1^{1/2}E_0$ for
the transverse field case. It is found that increasing $q$ causes
both the fundamental and third-order harmonics of the local
electric field to decrease. Accordingly, the fundamental harmonics
of the induced dipole moment is caused to decreasing as well.
However, the third-order harmonics of induced dipole moment is
caused to increase as $q$ increases. Note the three curves in
Fig.~1(c) are overlapped.

Similarly, we also investigate the harmonics of the local electric
field and induced dipole moment in Fig.~3, but for the
longitudinal field case. In contrast to the transverse field case
(Fig.~2), the exactly opposite effects have been shown for  the
longitudinal field case (Fig.~3), due to the opposite dependence
of $q$ on $\gamma'_x$ and $\gamma'_z ,$ see Fig.~1. Again, the three
curves in Fig.~1(c) are overlapped.

Both Figures~2~and~3 show that the harmonics are strongly dependent
on the nonlinear response of the suspended particles. Moreover,
increasing the nonlinear response $\chi_1^{1/2}E_0$ leads to an increase in the
harmonics. This is in agreement with the results of our recent
work~\cite{PRE1,PRE6}. In addition, we also studied the effect of volume fraction $\rho$, and found that increasing the volume fraction can enhance the harmonics too (no figures shown here).

To understand the results, we resort to the spectral
representation approach~\cite{Bergman}. Let's start by denoting
$s=(1-\ep_1/\ep_2)^{-1}$, and then the local electric field given
by Eq.~(\ref{ET}) admits \be \langle E_1\rangle
=E_0+\frac{F_1}{s-s_1}E_0, \ee where the residue $F_1=1/3$ and the
pole $s_1=(1-\rho\gamma')/3 .$ As $\rho\to 0 ,$ $s_1$ tends to $1/3
,$ which produces the known value for an isolated particle. The
substitution of the present model parameters yields $s=-0.081 ,$
and hence we have 
\be 
\langle E_1\rangle
=E_0+\frac{1/3}{-0.081-(1-0.1\gamma')/3}E_0. \label{SPR}
\ee 
Based on this
equation, it is apparent to see that increasing $\gamma'_x$ or
$\gamma'_z$ causes the electric field to decrease.
In addition, we have shown that increasing $q$ causes $\gamma_x'$ (or $\gamma_z'$) to increase (or decrease), see Fig.~1. Also, a plateau was shown at $\gamma_x'=\gamma_z'=1$, which actually includes the transformations ranging from the bcc ($q=1$) lattice to the fcc ($q=2^{1/3}$). Because of the dependence of the harmonics on $\gamma'$, similar plateau have already been shown in Figs.~2~and~3, too. 
 As a matter of fact, the effective linear dielectric constant mainly contributes to the magnitude of the fundamental harmonics of the induced dipole moment or local electric field. In the meantime, because of the existence of weak nonlinearity, the nonlinear polarization just has a perturbation effect, which can be neglected when compared to the linear part. Since the spectral representation approach is a different way of formulating the linear local electric field, to some extent, it can be used to explain the fundamental harmonics of the local electric field and hence those of the induced dipole moment which are displayed in Fig.~2(a,c) and Fig.~3(a,c). Based on Eq.~(\ref{SPR}) (namely, an expression obtained from the spectral representation approach), it is convenient for us to understand that a decrease (or increase) of $q$ should lead to an increase of the local electric field for transverse (or longitudinal) field cases. In this connection, for transverse (or longitudinal) field cases, the fundamental harmonics of the local electric field and hence corresponding induced dipole moment should increase as $q$ decreases (or increases). This has also been displayed in Fig.~2(a,c) (or Fig.~3(a,c)).
The explanation on the third-order harmonics of the local electric field and induced dipole moment seems to be more complicated. 
Nevertheless, since the strength of the
nonlinear polarization of nonlinear materials should be reflected in the magnitude of the high-order
harmonics of local electric fields and induced dipole moments, the effective nonlinear part in the effective nonlinear dielectric constant should be expected to determine the magnitude of the third-order harmonics. It appears that the third-order harmonics of the local electric field show a behavior similar to its fundamental harmonics. However, inverse behavior occurs to the third-order harmonics of the induced dipole moment. This difference would possibly be due to the nonlinearity fluctuation which might arise from the long-range interaction. This interaction existing in the system is explicitly evaluated in the current paper, by using the Ewald technique.

\section{Discussion and conclusion}

Here some comments are in order. In this work, we have used the
Ewald-Kornfeld formulation to derive the local electric fields and induced dipole moments in a rigorous manner, in an attempt to take into account the lattice
(local-field) effect on the nonlinear ac responses of ER solids under structure transformations.

In fact, we presented an (approximate) effective medium theory (EMT) for
considering the local-field effect on the electrorotation and
dielectric dispersion spectra of colloidal particles or biological
cells in the previous paper~\cite{PRE4}. However, we were unable
to study the detailed structural information via the EMT. In the
present paper, we have considered the local-field effect in a
rigorous manner, based on the Ewald-Kornfeld formulation which is
quite different from the EMT.

We have considered the fundamental and third-order harmonics. As a
matter of fact, it should be straightforward to investigate the
higher-order (e.g., fifth-order, seventh-order, etc.) harmonics
 by keeping more terms in Eq.~(\ref{Taylor-E}). Due to the virtue
 of symmetry, the odd-order harmonics appear only. In
 fact, the even-order harmonics can also be induced to occur  due to the
 coupling between the ac and dc external fields~\cite{WeiJAP}.
 Since the second-order harmonics can often be of several orders
 of magnitude larger than the third-order, it is of value to
 discuss the lattice effect on the nonlinear
 ac responses by applying an ac field coupling with a dc field,
in order to investigate the second-order harmonics. Finally,
this work can  be extended to polydisperse ER
solids~\cite{Sun03} in which the permittivities of particles can
possess a distribution, or to a system~\cite{HuangJPCB} where  anisotropies of material and geometry are caused to exist due to  fabrication methods or field effects. On the other hand,
the nonlinear response of composites to ac fields will become more complicated 
when the ac field contains different basic frequencies or different magnitudes of external 
fields at different basic frequencies~\cite{WeiJPD04}. 

Our former work~\cite{PRE1} covers a wider range of the degree of anisotropy $q$. However,
the explicit form of the local field-$q$ relation was lacking. By invoking
a lattice structure (which exists in ER solids), this can be achieved, as evident in this work.
In detail, in Ref.~\cite{PRE1} we used an effective medium theory to derive effective (anisotropic) dielectric constants, based on the Maxwell-Garnett approximation. In contrast, in the current paper, the Ewald technique is used to derive the effective dielectric constants, by including the long-range interaction explicitly (which was, however, lacking in Ref.~\cite{PRE1}).
In this sense, for treating ER solids with lattice structures, the present derivation of the effective dielectric constants should be expected to be more accurate.
Interestingly, some result of the current paper was also predicted in Ref.~\cite{PRE1}. For instance, as $q$ decreases, for longitudinal (or transverse) field cases $\gamma_z'$ (or $\gamma_x'$) is caused to increase (or decrease), thus yielding an increasing (or decreasing) third-order harmonics of the induced dipole moment, see Fig.~3(d) (or Fig.~2(d)). Here $\gamma_z'$ (or $\gamma_x'$) corresponds to $\beta_L$ (or $\beta_T$) of Ref.~\cite{PRE1}. It is worth to mentioning that a different normalization factor (i.e., $p_0$) was  used in Ref.~\cite{PRE1}.

%
%
%
%

To sum up, based on the Ewald-Kornfeld formulation, we have
investigated the nonlinear as responses of an ER solid with
nonlinear spherical particles embedded in a linear host fluid, and
found that the fundamental and third-order harmonic ac responses are
sensitive to the degree of anisotropy within the ER solid. Our
results have been well understood through the spectral
representation theory. Thus, by measuring the nonlinear ac responses of ER
solids, it seems possible to perform a real-time monitoring of 
structure transformations.

\section*{Acknowledgments}


This work was supported  by the Alexander von Humboldt foundation in Germany. The author acknowledges Professor K. W. Yu's very fruitful collaboration and comments.


 \newpage

\newpage
\begin{figure}[h]
\caption{Local field factor $\gamma'$ versus the degree of anisotropy $q .$
The bct (ground state), bcc and fcc lattices which are respectively related to
$q=0.87358$, $q=1.0$ and $q=2^{1/3}$
 are also shown (dot-dashed lines).  }
\end{figure}

\begin{figure}[h]
\caption{The fundamental and third-order harmonics of the local
electric field and induced dipole moment, as a function of $q$ for
$\chi_1^{1/2}E_0=0.7\ep_0^{1/2}$ (solid lines),
$\chi_1^{1/2}E_0=0.8\ep_0^{1/2}$ (dotted lines), and
$\chi_1^{1/2}E_0=0.9\ep_0^{1/2}$ (dashed lines),  for the
transverse field case. Note the three curves in (c) are
overlapped. Parameters: $\ep_1=30\ep_0 ,$ $\ep_2=2.25\ep_0 ,$ and
$\rho=0.1 .$ The bct, bcc and fcc lattices are shown as well
(dot-dashed lines).}
\end{figure}

\begin{figure}[h]
\caption{ Same as Fig.~2, but for the longitudinal field case.}
\end{figure}

\newpage
\centerline{\epsfig{file=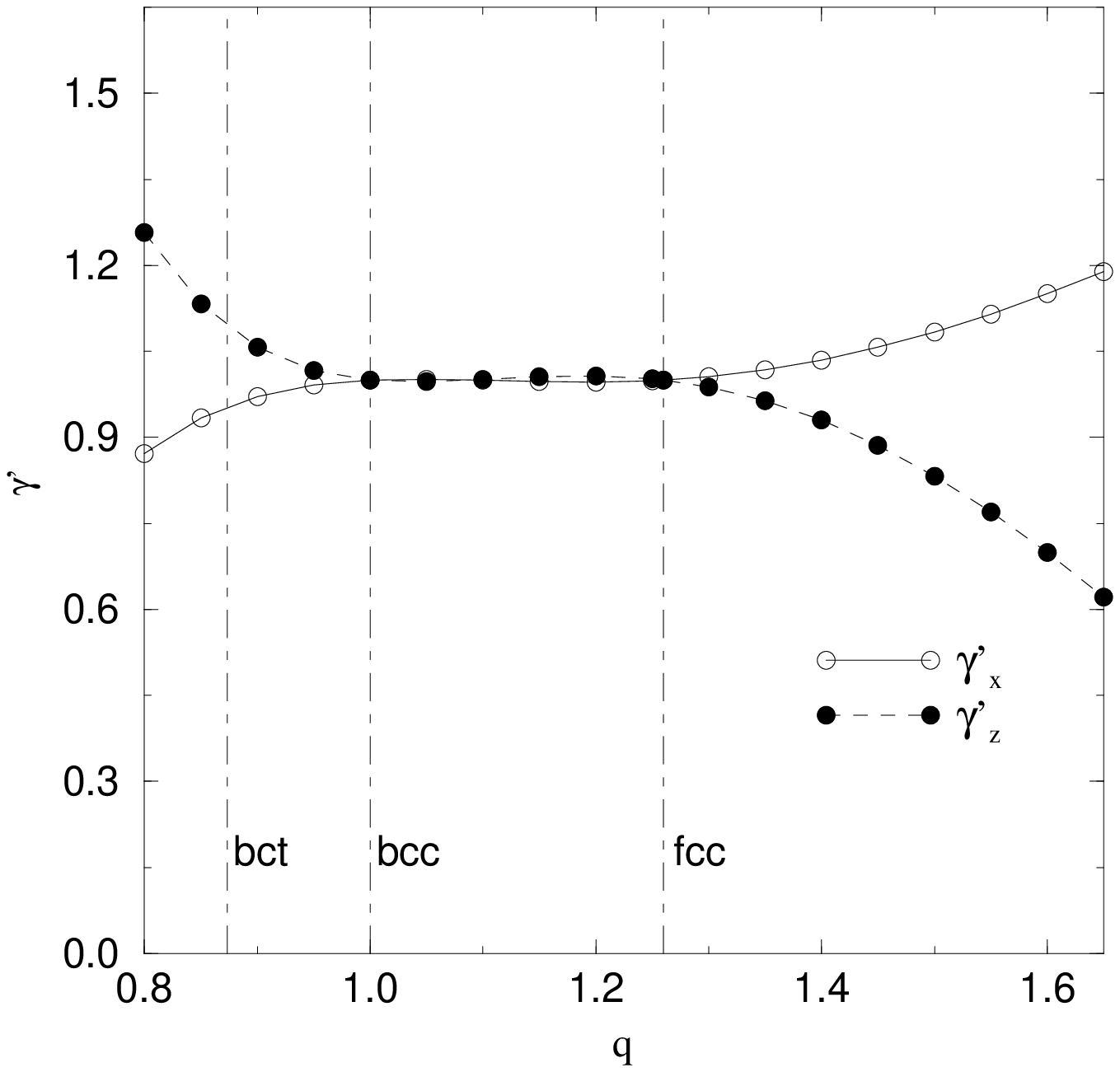,width=350pt}}
\centerline{Fig.~1/Huang}

\newpage
\centerline{\epsfig{file=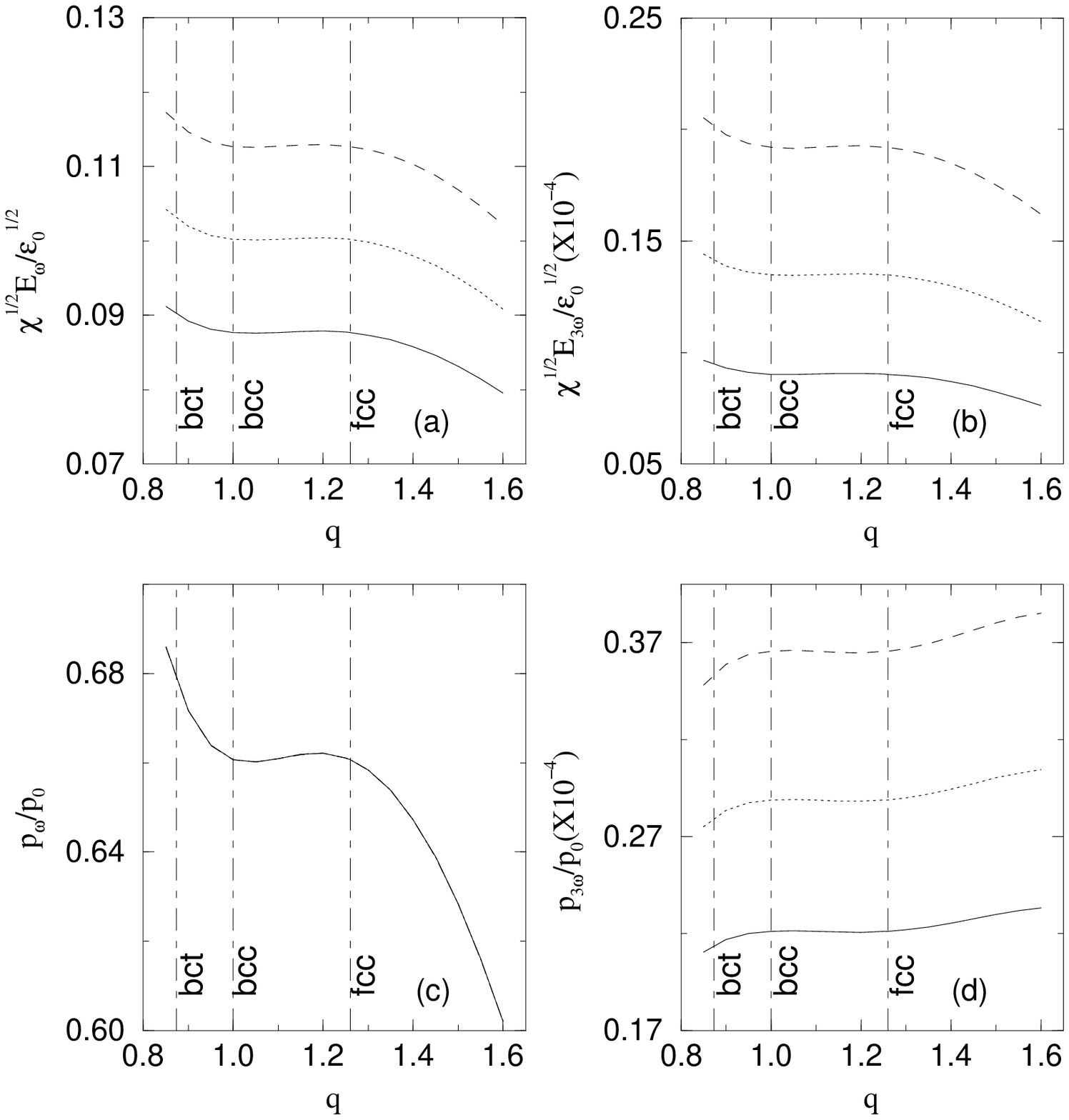,width=\linewidth}}
\centerline{Fig.~2/Huang}

\newpage
\centerline{\epsfig{file=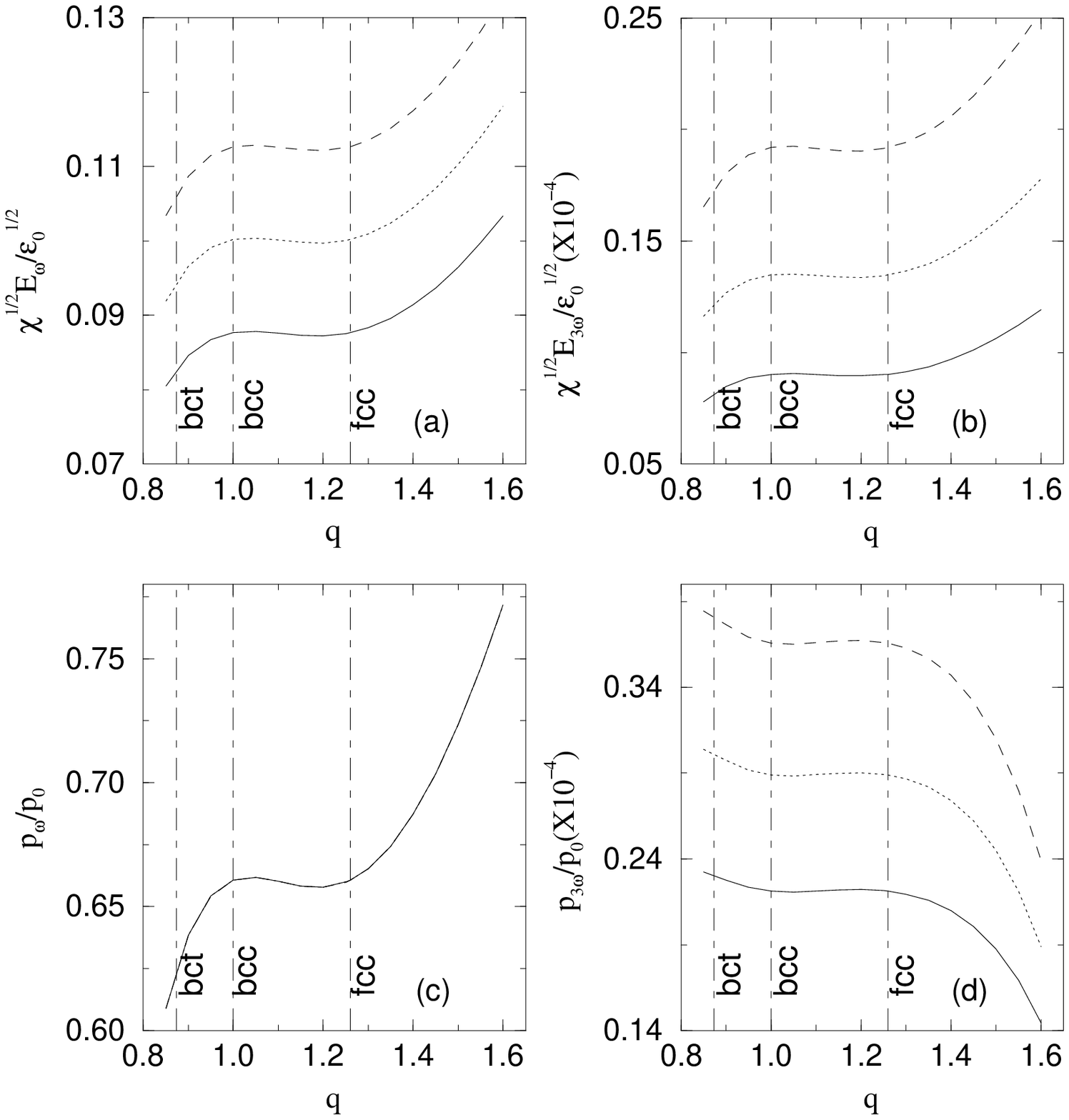,width=\linewidth}}
\centerline{Fig.~3/Huang}

\end{document}